\documentclass[a4paper,fleqn,usenatbib]{mnras}
  
\usepackage{newtxtext,newtxmath}

\usepackage[T1]{fontenc}
\usepackage{ae,aecompl}

\usepackage{graphicx}	
\usepackage{amsmath}	
\usepackage{amssymb}	

\usepackage{threeparttable}

\graphicspath{{figures/}}
 
\title[SZ and Dust in SHELA Clusters]{The Atacama Cosmology Telescope: SZ-based masses and dust emission from IR-selected cluster candidates in the SHELA survey}

\author[B. Fuzia et al.]{Brittany J. Fuzia,$^{1}$ 
Lalitwadee Kawinwanichakij,$^{2}$ 
Nicola Mehrtens,$^{2}$  
Simone Aiola,$^{3,4}$ 
\newauthor 
Nicholas Battaglia,$^{5}$ 
Robin Ciardullo,$^{6, 7}$ 
Mark Devlin,$^{8}$ 
Steven L.~Finkelstein,${^9}$ 
\newauthor 
Megan Gralla,$^{10}$ 
Matt Hilton,$^{11}$   
Kevin Huffenberger,$^{1}$\thanks{Email: khuffenberger@fsu.edu} 
John P. Hughes,$^{12}$ 
\newauthor 
Shardha Jogee,$^{9}$ 
Felipe A. Maldonado,$^1$ 
Lyman A.~Page,$^{3}$ 
Casey Papovich,$^{2}$ 
\newauthor 
Bruce Partridge,$^{13}$ 
Eli Rykoff,$^{14}$ 
Neelima Sehgal,$^{15}$
Crist\'obal Sif\'on,$^{16, 17}$ 
\newauthor 
Suzanne T.~Staggs,$^{3}$
Edward Wollack$^{18}$ \\
$^{1}$Department of Physics, Florida State University, Tallahassee, FL 32306, USA\\
$^{2}$Department of Physics and Astronomy, Texas A\&M University, College Station, TX 77843, USA\\
$^{3}$Department of Physics, Princeton University, Princeton, NJ 08544 \\
$^{4}$Center for Computational Astrophysics, Flatiron Institute, New York, NY 10003, USA\\
$^{5}$Department of Astronomy, Cornell University, Ithaca, NY~14853, USA \\
$^{6}$Department of Astronomy and Astrophysics, The Pennsylvania State University, University Park, PA 16802, USA\\
$^{7}$Institute for Gravitation and the Cosmos, The Pennsylvania State University, University Park, PA 16802, USA\\
$^{8}$Department of Physics and Astronomy, University of Pennsylvania, Philadelphia, Pennsylvania 19104, USA \\
$^{9}$Department of Astronomy, University of Texas at Austin, Austin, TX 78712 \\
$^{10}$Department of Astronomy/Steward Observatory, University of Arizona, Tucson, AZ  85721 \\
$^{11}$Astrophysics and Cosmology Research Unit, School of Mathematics, Statistics and Computer Science, University of KwaZulu--Natal,\\ \qquad Durban 4041, South Africa \\
$^{12}$Department of Physics and Astronomy, Rutgers University, 136 Frelinghuysen Road, Piscataway, NJ 08854 \\
$^{13}$Department of Physics and Astronomy, Haverford College, Haverford, PA 19041 \\
$^{14}$ SLAC National Accelerator Laboratory, Menlo Park, CA 94025, USA \\
$^{15}$Physics and Astronomy Department, Stony Brook University, Stony Brook, NY 11794 \\
$^{16}$Instituto de F\'isica, Pontificia Universidad Cat\'olica de Valpara\'iso, Casilla 4059, Valpara\'iso, Chile \\
$^{17}$Department of Astrophysical Sciences, Princeton University, Princeton, NJ 08544 \\
$^{18}$NASA/Goddard Space Flight Center, Greenbelt, MD 20771 \\
}

\date{Accepted XXX. Received YYY; in original form ZZZ}

\pubyear{2019}

\begin{document}
\label{firstpage}
\pagerange{\pageref{firstpage}--\pageref{lastpage}}
\maketitle

\begin{abstract}
We examine the stacked thermal Sunyaev-Zel\text{'}dovich (SZ) signals for a sample of galaxy cluster candidates from the Spitzer-HETDEX Exploratory Large Area (SHELA) Survey, which are identified in combined optical and infrared SHELA data using the redMaPPer algorithm. 
We separate the clusters into three richness bins, with average photometric redshifts ranging from 0.70 to 0.80. 
The richest bin shows a clear temperature decrement at 148 GHz in the Atacama Cosmology Telescope data, which we attribute to the SZ effect.  
All richness bins show an increment at 220 GHz, which we attribute to dust emission from cluster galaxies. 
We correct for dust emission using stacked profiles from Herschel Stripe 82 data, and allow for synchrotron emission using stacked profiles created by binning source fluxes from NVSS data.
We see dust emission in all three richness bins, but can only confidently detect the SZ decrement in the highest richness bin, finding $M_{500}$ = $8.7^{+1.7}_{-1.3} \times 10^{13} M_\odot$.  
Neglecting the correction for dust depresses the inferred mass by 26 percent, indicating a partial fill-in of the SZ decrement from thermal dust and synchrotron emission by the cluster member galaxies.
We compare our corrected SZ masses to two redMaPPer mass--richness scaling relations and find that the SZ mass is lower than predicted by the richness.
We discuss possible explanations for this discrepancy, and note that the SHELA richnesses may differ from previous richness measurements due to the inclusion of IR data in redMaPPer.
\end{abstract}




\section{Introduction}
Clusters of galaxies are the largest gravitationally bound structures in the universe. 
They are powerful cosmological probes because they sample the maxima of the primordial density field, and allow us to gain insight into large-scale structure, galaxy evolution, dark matter dynamics, and cosmological parameters  \citep{1999PhR...310...97B, 2002ARA&A..40..643C, 2005RvMP...77..207V}. 
The thermal Sunyaev-Zel'dovich (SZ) spectral distortion of the Cosmic Microwave Background (CMB) \citep{1972CoASP...4..173S} can be used as an indirect measurement of one of the most important observables---total galaxy cluster mass---and can identify clusters to high redshift \citep{2018ApJS..235...20H, 2018AstL...44..297B, 2019ApJ...870....7K}. 
The SZ effect occurs when CMB photons scatter from the hot electron gas in the intracluster medium. 
Only a small fraction of the photons interact: a $10^{14}\ M_{\sun}$ cluster has a $1.3'$-aperture-averaged optical depth of $\tau \sim 2\times 10^{-3}$ \citep{2016JCAP...08..058B}.  The photons gain energy through inverse Compton scattering, which alters the observed CMB spectrum and results in a characteristic spectral dependence for the SZ effect: a flux decrement for frequencies below 217 GHz and a flux increment for higher frequencies \citep{1972CoASP...4..173S}.
The magnitude of the effect is proportional to the Comptonization parameter (the integrated electron pressure), and the pressure is proportional to the depth of the gravitational potential well.  Therefore the amplitude of the SZ signal depends closely, but not linearly, on the mass of the cluster.
Using the SZ effect for cosmology requires an understanding of this relationship between the halo mass and the SZ observable, which is often expressed as the Comptonization parameter integrated over the cluster's solid angle. 
Since the SZ effect is a distortion of the CMB's spectrum, the signal does not decrease with distance the way that the cluster emission does, and so the SZ effect is an efficient way to find high-redshift clusters, limited only by the mass of the cluster and the sensitivity of the telescope.

At high mass, $M \gtrsim 10^{15}$ $M_\odot$, current SZ searches can already find halos efficiently at all redshifts, but at lower mass, $M \lesssim 10^{14}$ M$_{\odot}$, it becomes more difficult. 
These lower mass halos are interesting because their smaller potential wells have a harder time holding onto their gas, and are laboratories for star formation and AGN feedback \citep{2011MNRAS.412..947B,2013A&A...555A..66L, 2015MNRAS.448.2085L}.  
Although studies of low mass halos using the SZ effect will become more common as CMB telescopes become more sensitive, for now we depend on stacking, or averaging, multiple clusters that have been detected by other means.  
Spatially coherent stacking allows the use of the SZ effect to extend to lower masses, as it averages out contributions from the CMB, atmosphere, and detector noise \citep{2017MNRAS.468.3347S,2013ApJ...767...38S, 2011A&A...536A..12P, 2011ApJ...736...39H}.

At low redshift, optical surveys can identify clusters efficiently with multiband overdensity finders: e.g., MaxBCG \citep{2007ApJ...660..221K}, redMaPPer \citep{2014ApJ...785..104R}, and CAMIRA \citep{2014MNRAS.444..147O}.
At higher redshift, the infrared becomes an efficient avenue of detection: e.g., MaDCoWS \citep{2014ApJS..213...25S,2019ApJS..240...33G}, ISCS 
\citep{2008ApJ...684..905E}, IDCS \citep{2012ApJ...753..164S}, RCS \citep{2000AJ....120.2148G}, and several Spitzer catalogs \citep{2008ApJ...676..206P, 2010ApJ...716.1503P}.

The SZ signals of low-richness, optically-selected clusters are smaller than expected from mass--richness relationships, which are usually calibrated with high-richness clusters \citep{2011A&A...536A..12P,2012PhRvD..85b3005D,2013ApJ...767...38S,2017MNRAS.468.3347S}. 
Several possible explanations for this discrepancy are: radio or infrared point source contamination of the SZ signal, line-of-sight projections contaminating richness measurements, cluster miscentering, variable gas mass fractions in optically selected clusters, or more fundamentally, a lower amplitude for the mass-richness relation.
Solving this discrepancy is vital so that scaling relations for clusters, and therefore cluster physics, are understood over a wide mass range, allowing clusters to be used to their full cosmological potential.

In this work, we look for the SZ signal in data from the Atacama Cosmology Telescope (ACT), using cluster candidates selected by the redMaPPer algorithm from catalogs of multiwavelength imaging, including Spitzer data from the SHELA survey.
The resulting sample is higher in redshift and lower in mass than many other samples. 
Using Herschel and NRAO VLA Sky Survey (NVSS) data, we correct the stacked SZ decrement for contamination from dust and synchrotron emission, while simultaneously fitting for a halo mass based on the stacked SZ signal.
We characterize the uncertainty with a Markov Chain Monte Carlo.
We also compare the sample's SZ masses to optical mass-richness relationships.

We adopt a flat $\Lambda$CDM cosmology with parameters from  \cite{2016A&A...594A..13P}: $H_{0}$ = 67.3 km s$^{-1}$ Mpc$^{-1}$ and $\Omega_{\rm m}$ = 0.315. 
The  mass $M_{500}$ is measured out to $R_{500}$, which is the radius enclosing 500 times the critical density at a given redshift.

This paper is organized as follows: In Section \ref{sec:data} we introduce the cluster sample, the ACT and ACTPol data, the Herschel data, and the NVSS data used in this analysis. 
In Section \ref{sec:methods} we describe the methods we used to analyze the data. 
These include the filtering and stacking procedures, calculation of the covariance matrices, and a discussion of the noise and signals that contribute to the stacked profiles. 
We describe our resulting multifrequency stacked profiles 
and discuss our methods for removing dust and synchrotron contamination. 
We describe our fitting procedure, including the SZ and pressure profile we use to translate our SZ signal into a cluster mass. 
In Section \ref{sec:results}, we present the results of our analysis and discuss how our SZ masses scale with richness. 
In Section \ref{sec:conclusions}, we conclude with a summary of the analysis and results.

\section{Data} \label{sec:data}

\subsection{Cluster Sample}

\begin{table}
  \centering
  \caption{Properties of the SHELA cluster candidate richness bins}
  \begin{tabular}{|*{4}{c|}}
    \hline
    & $10 \leq \lambda < 20$ & $20 \leq \lambda < 30$ & $\lambda \geq 30$ \\ \hline
    
    $N_{\rm clusters}$ & 840 & 172 & 70 \\ \hline
     
    $z$ range & 0.50--1.60 & 0.50--1.35 & 0.52--1.18  \\ \hline
    
    average $z$ & 0.80 & 0.73 & 0.70 \\ \hline
    
    $\lambda$ range & 10--20 & 20--30 & 30--76 \\ \hline 
    
    average $\lambda$ & 14 & 24 & 39 \\ \hline
    
    \end{tabular}
\label{table:sampleinfo}
\end{table}

The sample contains IR- and optically-selected redMaPPer cluster candidates from the Spitzer-HETDEX Exploratory Large Area survey  \citep[SHELA]{2016ApJS..224...28P}. 
SHELA is a 24 deg$^2$ IRAC 3.6 and 4.5 micron survey in a low IR background region of Stripe 82 \citep{2000AJ....120.1579Y}, centered at a right ascension of 1$^{\rm h}$22$^{\rm m}$00$^{\rm s}$ on the celestial equator, and extending $\pm$6.5$^{\circ}$ in right ascension and $\pm$1.25$^{\circ}$ in declination. 
The SHELA survey region also includes DECam $ugriz$ imaging. Multiwavelength coverage in the same field includes SDSS and HETDEX in the optical, NEWFIRM in K-band, Herschel in the sub-mm, and ACT in the microwave. 

For this study, we use a galaxy catalog based on DECam and SHELA imaging \citep{2019ApJS..240....5W}.  We process the galaxy catalog with the redMaPPer algorithm \citep{2014ApJ...785..104R}, resulting in a catalog of 1082 groups and clusters with a richness $\lambda \geq 10.$ Richness is a measure of how many galaxies belong to a cluster. 
In redMaPPer, it is defined as the sum of the membership probabilities for the galaxies within a cluster. 

We use clusters with richnesses $\lambda \geq 10$, and break these into three richness bins: $10 \leq \lambda < 20$, $20 \leq \lambda < 30$, and $\lambda \geq 30$. 
{There are 840, 172, and 70 clusters in the lowest- to highest-richness bins, respectively.} 
Two rich clusters from the SHELA sample have already been detected in the ACT SZ cluster sample in this area of the sky.
The ACT-detected clusters are ACT-CL J0058.0+0030 with a S/N of 5.0 and ACT-CL J0059.1-0049 with a S/N of 8.4 \citep{2013JCAP...07..008H}.  
None of the remaining objects are detected individually in SZ by ACT, so their individual masses must be roughly $\leq$ $10^{14}$ M$_\odot$, ACT's approximate mass limit.

In the relevant redshift range, $z \sim 0.7$--$0.8$, the 90\% completeness limit for $M_{500}$ in ACT is $4$--$5 \times 10^{14}$ M$_{\odot}$ \citep{2018ApJS..235...20H}.   
Properties of each richness bin are summarized in Table \ref{table:sampleinfo}.

\subsection{ACT Millimeter-Wave Data}
We use ACT data to measure the SZ decrement and null signals. 
ACT is a six-meter millimeter-wave telescope at an altitude of 5200 meters on Cerro Toco in the Chilean Atacama Desert \citep{2011ApJS..194...41S}. 
It surveys the CMB with high resolution and sensitivity. 
The first generation of ACT observations dates from 2007-2010; there were three detector arrays operating at frequencies of 148, 220, and 277 GHz. 
These bands were chosen to study the SZ and capture the SZ decrement, null, and increment. 
ACT surveyed two regions on the sky, the ``southern'' and ``equatorial'' surveys. 
The southern survey covered 455 deg$^2$ and is centered on declination -53.5$^{\circ}$ \citep{2011ApJ...731..100M}. 
The equatorial survey overlaps with 270 deg$^2$ of Stripe 82 and the entire SHELA survey, covering 504 deg$^2$ and spanning from 20$^{\rm h}$16$^{\rm m}$00$^{\rm s}$ to 3$^{\rm h}$52$^{\rm m}$24$^{\rm s}$ in right ascension and $-2^{\circ}$07$^{\prime}$ to 2$^{\circ}$18$^{\prime}$ in declination \citep{2013JCAP...07..008H}. 
The second generation of the experiment, ACTPol, was deployed in 2013   \citep{2016ApJS..227...21T}.
It has receivers at 90 and 148 GHz, polarization capability,  and triple the sensitivity of ACT. 
ACTPol has made observations in four deep field patches and one wider field \citep{2014JCAP...10..007N,2018ApJS..235...20H}. 
The wider ``D56'' region overlaps with SHELA, covers 548 deg$^2$, is centered on the celestial equator, and expands the area covered by ACT.\footnote{ACT and ACTPol maps are available for download from \url{https://lambda.gsfc.nasa.gov/product/act/}}
In this work, we measure the SZ decrement using coadded, point source subtracted, ACT temperature maps at 148 GHz from all observing seasons that overlap with the SHELA survey region: seasons 3--4 (2009 and 2010) of ACT and season 2 (2014) of ACTPol. 
The ACT maps have 0.495 arcmin pixels, while the ACTPol maps have 0.5 arcmin pixels. 
We use ACT maps that have been repixelized into the ACTPol pixelization to make coadded maps of all available data. 
From seasons 3--4 of ACT, we use data at 220 GHz, a frequency near the SZ null, to constrain contamination from thermal dust and synchrotron emission. 
These maps are also repixelized into the ACTPol pixelization. 
At FWHM, the beam sizes are 1.4 arcmin at 148 GHz, and 1.0 arcmin at 220 GHz.

\subsection{Herschel Submillimeter Data}
To measure dust emission from cluster member galaxies, we use far-IR data from the Herschel Stripe 82 (HerS) survey, which consists of maps at 250, 350, and 500 $\mu$m (or 1200, 857, 600 GHz) observed with \textit{Herschel}/SPIRE \citep{2014ApJS..210...22V}. 
The survey covers 79 deg$^2$, spanning 13$^{\circ}$ to 37$^{\circ}$ in right ascension and -2$^{\circ}$ to +2$^{\circ}$ in declination. 
The SPIRE beams are 18.2, 25.2, and 36.3 arcsec at 1200, 857, 600 GHz, respectively. 
In addition to the maps, in this work we use the band-merged source catalog from the HerS team, which contains compact source flux densities and uncertainties in each band. 
The HerS team assumed sources are point-like and they identified them using the IDL software package STARFINDER \citep{2000A&AS..147..335D}. 
They produced the band-merged catalog using the De-blended SPIRE Photometry (DESPHOT) algorithm, which uses source positions from the 1200 GHz band as a prior for the other frequencies \citep{2010MNRAS.409...48R}.

\subsection{NVSS Radio Data}
To measure synchrotron emission from cluster member galaxies, we use 1.4 GHz data from the NRAO VLA Sky Survey (NVSS), which covers the sky North of declination -40$^{\circ}$ \citep{1998AJ....115.1693C}.
We use flux densities and uncertainties from the source catalog, which contains around 10$^6$ sources brighter than approximately 2.5 mJy.
From the lowest to highest richness bin respectively, there are 5428, 1127, and 485 total sources within 9$^\prime$-radius apertures centered on the clusters.

\section{Methods} \label{sec:methods}

Our overall strategy in this analysis is to build up a model for the stacked emission in the ACT bands at the location of the clusters, allowing for SZ, dust, and synchrotron components.  
We estimate the data's covariance due to CMB and noise fluctuations, and then use a Markov Chain Monte Carlo to estimate the parameters of our emission model.

\begin{figure*}
\centering
  \includegraphics[width=1.75\columnwidth]{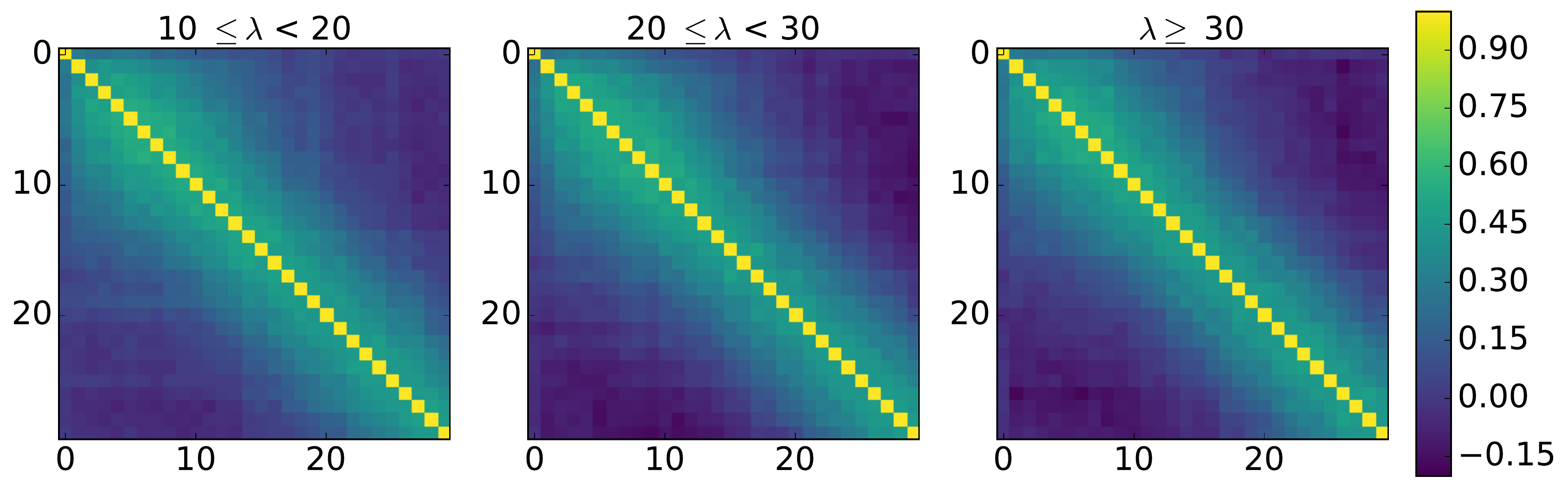}
  \caption{Correlation matrices for the stacked pressure profiles at 148 GHz for the three richness bins, $10 \leq \lambda < 20$ (left), $20 \leq \lambda < 30$ (middle), and $\lambda \geq 30$ (right). They are obtained by stacking on 1600 simulations of ACT which contain correlations introduced by CMB fluctuations, detector noise, and atmospheric noise, and account for the different observing seasons of ACT used in this analysis. Adjacent bins are correlated at $\sim$0.5. The axes label the radial bin numbers, which extend to $9^\prime$.}
  \label{fig:covariance}
\end{figure*}

\subsection{SZ Profiles}
The SZ signal can be expressed as a change in CMB temperature by: 
\begin{equation}
\centering
  \frac{\Delta T}{T_{\rm CMB}} = y( \theta ) \, f( x ) ,
\end{equation}
where $x = {h \nu}/{k T_{\rm CMB}}$ and $f(x) = x ({e^{x} + 1})/{(e^{x}-1)} - 4$ contains the frequency dependence of the SZ effect in the non-relativistic limit. The Compton parameter $y$ is proportional to pressure integrated along the line of sight \citep{1972CoASP...4..173S,1970CoASP...2...66S}:
\begin{equation}
  y(\theta) = \frac{\sigma_{T}}{m_{e} c^{2}} \int dl\  P(\theta, l) . \end{equation}
To translate our stacked temperature profile into a mass estimate, we use the universal pressure profile (UPP) of \cite{2010A&A...517A..92A} [A10], which is calibrated using low-redshift X-ray clusters from REXCESS \citep{2007A&A...469..363B}. 
A10 fit a generalized Navarro-Frenk-White profile which allows for a normalization that varies with mass and redshift and a mass-dependent deviation from self-similarity in the shape of the profile. 
In this model, the pressure at any radius $r$ (or $x' \equiv r/R_{500}$) is:

\begin{equation}
P(r) = P_{500} \bigg[ \frac{M_{\rm SZ,500}}{3x10^{14} h^{-1}_{70} M_{\odot}} \bigg]^{\alpha_{p} + \alpha^{\prime}_{p}(x')} \textbf{p}(x') \ h^{2}_{70} \ \mbox{keV} \ \mbox{cm}^{-3},
\end{equation}
where $P_{500}$ is the normalization of the pressure profile at the radius where the density is 500 times the critical density at a given redshift,
\begin{equation}
P_{500} = 1.65 \times 10^{-3} E(z)^{8/3} \bigg[\frac{M_{500}}{3 \times 10^{14} h_{70}^{-1} M_{\odot}}\bigg]^{2/3} h_{70}^2\ \mbox{keV cm}^{-3},
\end{equation}
$E(z)$ is the evolution of the Hubble parameter, \textbf{p}$(x')$ is the dimensionless universal pressure profile,
\begin{equation}
\textbf{p}(x') = \frac{P_{0}}{(c_{500} x')^{\gamma} [1 + (c_{500} x')^{\alpha}]^{(\beta-\gamma)/\alpha}},
\end{equation}
and $\alpha^{\prime}_{p}(x')$ describes the deviation from self-similiarity,
\begin{equation}
\alpha^{\prime}_{p}(x') = 0.10 - (\alpha_{p} + 0.10) \frac{(x'/0.5)^{3}}{1 + (x'/0.5)^{3}},
\end{equation}
with $\alpha_p$ = 0.12.
Using local clusters with XMM-Newton data, \cite{2013A&A...550A.131P} update the best-fit parameters for the UPP to [$P_{0},c_{500},\gamma,\alpha,\beta$] = [6.51,1.81,0.31,1.33,4.13], which we adopt in this paper.

\subsection{Map Filtering and Stacking}
Before stacking, we filter the maps using a  high-pass filter designed to lessen the impact of large scale CMB fluctuations while minimally altering the small scale cluster signal. 
To avoid bias, we design a filter independent of any assumed cluster shape.
Our filter is a Fourier-space high-pass filter that we define in terms of its low-pass complement.  
The low-pass complement in real space is an apodized top-hat. 
It is unity inside $3^\prime$ radius and tapers to zero outside of $5^\prime$ with a cosine transition. 
Thus our high-pass filter removes the large scale features in the map, and barely touches small harmonic scales. 
Our filter is not matched to any specific cluster profile, and leaves much of the small-scale detector white noise in the data.  
A less aggressive filter, which tapers to zero between $9^\prime$ and $11^\prime$ produces profiles consistent with those shown in Figure \ref{fig:rawstacks}.
When compared, all maps, simulations, and model cluster profiles are subjected to this same filter. 

Our results are robust to changes in this filter.  As a test, we modified the filter to introduce beam smoothing.  The filter is the same high-pass filter used for the analysis, but convolved with the beam at 148 GHz to reduce pixel-scale white noise. 
This resulted in smoother, shallower profiles, but the mass estimates were similar to those reported in this analysis, with higher uncertainty as there was more bin to bin correlation.

Most of the cluster candidates were not individually detected in SZ by ACT. 
To increase the signal-to-noise we stack, or average, observations of the clusters together into 30 annular bins, centered on the redMaPPer cluster positions, out to a radial separation of 9$^\prime$, which is chosen to be past the filtering scale. 
CMB and white noise fluctuations have zero mean, therefore stacking observations partially averages out these noise fluctuations. 
Measurements for a given pixel are placed in the radial bin in which the center of that pixel falls.

We can write our stacked profiles as a sum of the beam-convolved signals $(b \ast P)(\theta)$, noise ($n$), and a DC offset ($p_0$):
\begin{equation}
\begin{split}
      P^{148} & = b^{148} \ast ( P^{\rm SZ} + P^{\rm dust,148} + P^{\rm synch,148} + n^{\rm CMB} ) \\ & + n^{\rm det,148} + p_0^{148}.
\end{split}
\label{eq:prof148}
\end{equation}

\begin{equation}
  \label{eq:prof220}
 \begin{split}
  P^{220} & = b^{220} \ast ( P^{\rm dust,220} + P^{\rm synch,220} + n^{\rm CMB} ) \\ & + n^{\rm det,220} + p_0^{220}.
\end{split}
\end{equation}

The ACT beams ($b^{148}, b^{220}$) include the smoothing effects of the pixel window and the telescope pointing jitter.
There is little SZ signal at 220 GHz, so we can use this profile to estimate the contributions from dust and synchrotron emission in the ACT bands. 
The models we use to estimate $P^{\rm dust}$ and $P^{\rm synch}$ are discussed in Sections \ref{sec:dustprof} and \ref{sec:radioprof}.

\subsection{Covariance Matrices}

In the ACT frequencies, 148 and 220 GHz, the error in each annular bin of the stacked profile reflects the covariance introduced by CMB fluctuations, detector noise, and atmosphere. 
The covariance matrix is calculated by carrying out the same filtering and stacking procedure on 1600 simulations that model coadded ACT and ACTPol maps.
As a first step, we make a coadded data map by repixelizing the ACT maps into the new ACTPol pixelization, and sum up all the different seasons and arrays. 
We use the power spectra of that coadded data to generate CMB plus noise simulations.  These account for cross-correlations between 148 and 220 GHz present in the data (chiefly due to the CMB, but also correlated atmosphere).  
The mock maps are then filtered the same way as the data prior to stacking. 
After filtering, and for each richness bin, we stack on each of the 1600 simulations at the actual locations of the SHELA clusters.  Using the actual locations helps to capture the correct correlations from the random realizations of CMB and noise.
We use the stacked profiles to compute the covariance between the annular bins.
The covariance is largest in the small angle bins where there are few measurements to average down the noise.
We compute the correlation matrix from the covariance matrix (which normalizes the diagonal) and show it in Figure \ref{fig:covariance} for each richness bin at 148 GHz.  There is mild correlation between the annular bins.  

Errors on the dust SED, dust profiles, and synchrotron profiles also figure into our final uncertainties.  
The HerS catalog includes 1$\sigma$ uncertainties for the flux density of each source, which we use by averaging in quadrature for the SED fitting process. 
The error bars on the stacked Herschel profiles come from stacking on the survey's noise maps. 
The covariances of the stacked NVSS profiles come from binning the variance of each source flux and smoothing with the appropriate ACT beam and filter to account for the bin-to-bin correlation.

\subsection{Multifrequency Profiles} \label{sec:multifreqprof}

\begin{figure*}
  \centering
  \includegraphics[height=0.85 \textheight]{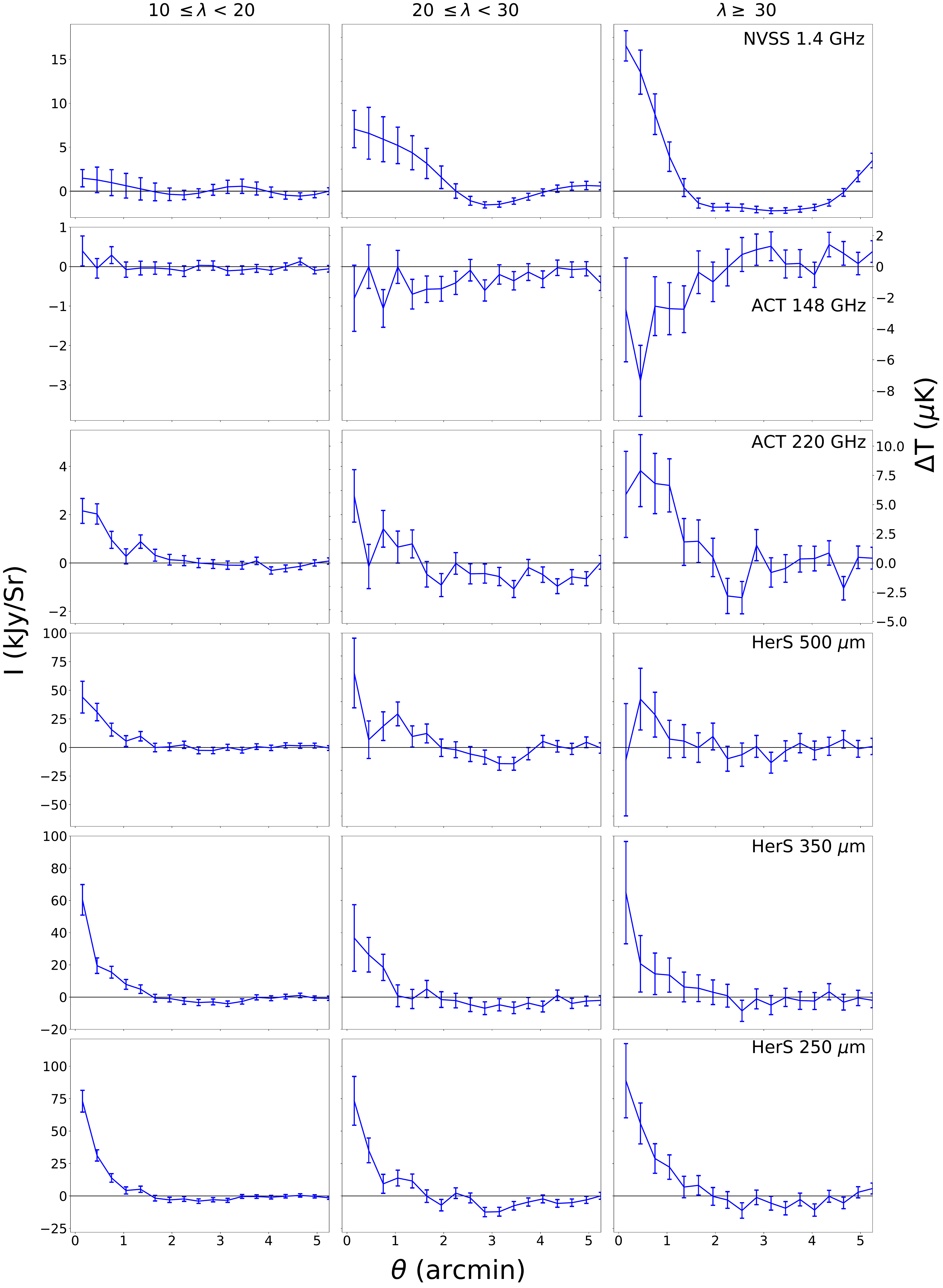}

\caption{Stacked profiles of the three richness bins for all 6 bands used in this analysis: NVSS at 1.4 GHz, ACT at 148 GHz and 220 GHz, and HerS at 500 $\mu$m, 350 $\mu$m, and 250 $\mu$m. 
The main contributions to the profile at 148 GHz are the SZ signal and dust and synchrotron emission. 
ACT 220 GHz is near the SZ null, and shows emission which also contaminates the signal at 148 GHz. 
The Herschel bands (500, 350, 250 $\rm \mu m$) trace thermal dust emission in the clusters, and the profiles are the result of stacking on each frequency map and subtracting DC offsets from radii > 5$^\prime$. 
The NVSS stacks trace synchrotron emission, and the profiles result from binning NVSS sources based on their angular separation from the cluster centers and smoothing with the ACT 148 GHz beam. 
The NVSS stacks sometimes go negative due to the hi-pass filtering, and the bins are highly covariant due to the beam smoothing.
The error bars are the diagonal of the covariances described in the text.
All profiles are filtered with the high-pass filter.
We plot our stacked profiles out to a radius of 5$^\prime$ to highlight the cluster-centered emission.}  
  \label{fig:rawstacks}
\end{figure*}

\begin{figure}
  \centering
  \includegraphics[width=0.45\textwidth]{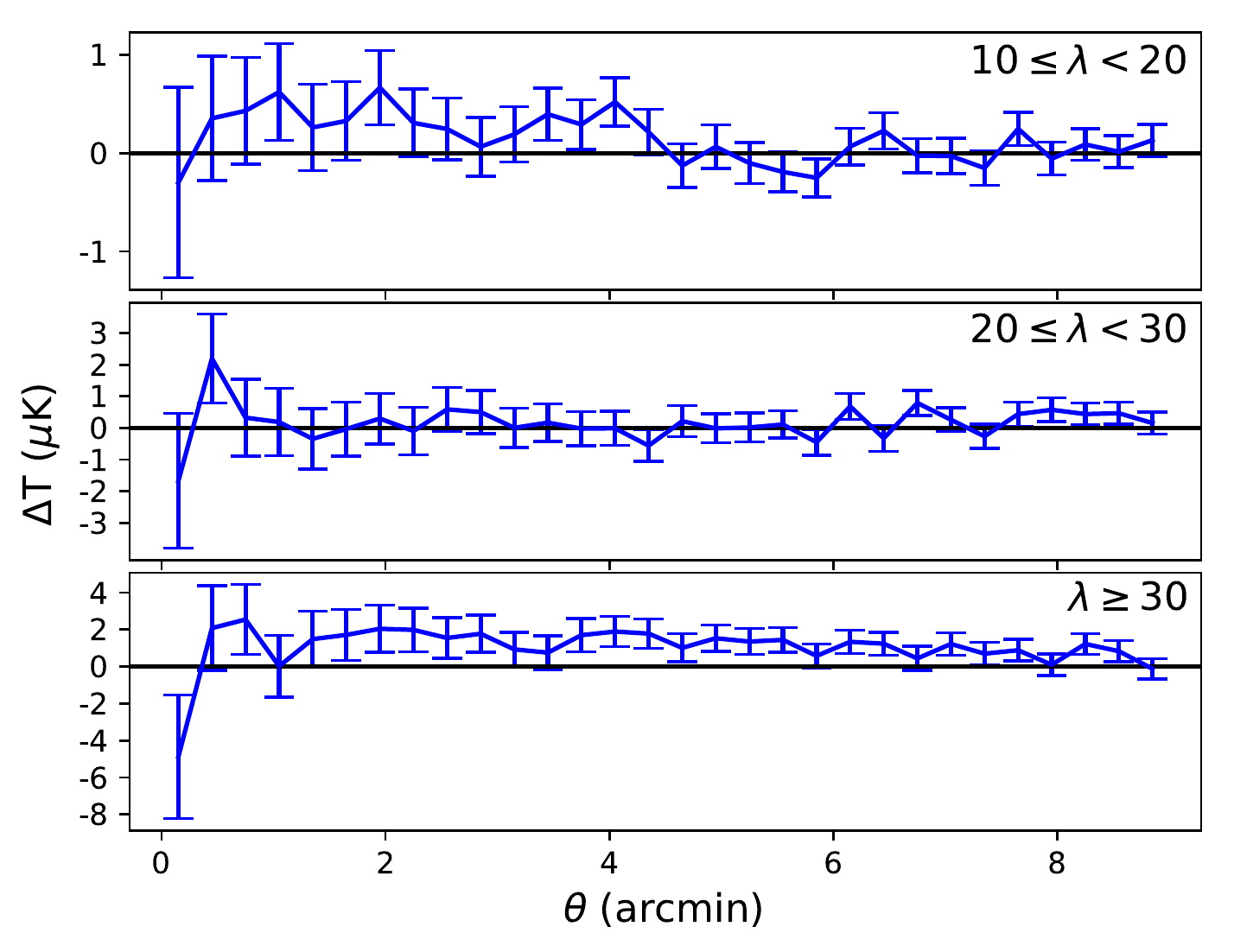}

\caption{ Stacking null test at 148 GHz. We generate random positions for the number of clusters in each richness bin and stack the ACT map in those locations to test that there is no signal in places unassociated with clusters.  Although the bottom bin has an offset, it has the smallest number of objects and so is most subject to large scale fluctuations.  Assessed by the $\chi^2$, none of these random samples have a significant signal, as expected. The bins are highly correlated as shown in Fig \ref{fig:covariance}.}  
  \label{fig:randstacks}
\end{figure}

The stacked profiles for the different richness bins are shown in Figure \ref{fig:rawstacks}. 
The profiles show the cluster-centered emission for NVSS sources at 1.4 GHz, ACT at 148 and 220 GHz, and the three HerS bands (600--1200 GHz).\footnote{When the NVSS synchrotron profile is extrapolated to the ACT bands, it is convolved with the appropriate ACT beam, making it much smoother.}
Here, as an example, we plot the synchrotron profile and error bars smoothed with the 148 GHz beam.
The ACT profiles at 148 GHz do not show an SZ decrement at high significance, but any decrement is subject to being filled in by the dust emission and synchrotron emission, which are two mechanisms we want to constrain.
We expect the stacks at 220 GHz to contain little SZ signal, but they do show clear cluster-centered emission. 
We use this 220 GHz emission profile to constrain the dust and synchrotron components. 
We also show profiles from the HerS maps at 500, 350, and 250 $\mu$m that we use to fit for dust. 
We have fit and removed offsets in the profiles at radii larger than 5$^\prime$.
The error bars on the stacked HerS profiles result from stacking on the noise maps provided with the HerS data. 

The ACT 220 GHz, NVSS, and Herschel stacks demonstrate that there is a signal from dust and synchrotron emission within a few arcminute radius of the cluster centers:
compared to the null hypothesis of zero emission, the probabilities to exceed $\chi^2$ for the ACT 220 GHz profiles and the nine Herschel profiles shown in Figure \ref{fig:rawstacks} are each < 0.05, so we conclude that cluster-centered emission is present.
For the NVSS profiles, we calculate the probability to exceed $\chi^2$ before they are smoothed with the ACT 148 beam, and find values which are also < 0.05.  (Figure \ref{fig:rawstacks} shows them after smoothing.)
The probability to exceed $\chi^2$ for the ACT 148 profiles are 0.69, 0.35, and 0.03, respectively for the lowest to highest richness bins.

To test the robustness of the noise model, and that our signals are not merely a feature caused by the stacking procedure, we stack on random positions in the ACT maps, and find no signal on average. 
Figure \ref{fig:randstacks} shows the results from stacking on random positions in the 148 GHz map, with each null stack accounting for the number of clusters in the different richness bins.   
We use the full covariance to compute $\chi^2$.  
The probabilities to exceed $\chi^2$ for the random position stacks are 0.47, 0.45, and 0.35 respectively for the lowest to highest richness bin, consistent with no emission in the null stacks.  
When stacking on the same random positions in the 220 GHz map, the probabilities to exceed $\chi^2$ are 0.12, 0.35, 0.70, from the lowest to highest richness bin.

\subsection{Dusty Source Contamination}
\label{sec:dustprof}
Emission from dusty sources and radio sources contaminates the SZ signal in clusters \citep{2005A&A...439..901A}. As seen in Figure \ref{fig:rawstacks}, when stacking at 220 GHz (the SZ null) and higher frequencies, there is an excess signal, which we partially attribute to dust emission from cluster member galaxies. 
This positive emission will also be present at 148 GHz, where it fills in the SZ decrement, causing the sample to appear to have less than its true mass. 
To correct for this, we fit for a dust component at 148 and 220 GHz.  

The dust profiles at 148 and 220 GHz take their shape from the HerS stacked profiles.  
Using a dust SED that we fit to HerS sources, we have extrapolated the three HerS profiles to the ACT frequencies, and then averaged them with inverse variance weighting.
The unnormalized dust profile $P^{\rm dust}$ in the ACT bands $\nu_{\rm ACT}$ for bin $b$ is given by:
\begin{equation}
    P^{\rm dust,\nu_{\rm ACT}}_b = \frac{ \sum_{\nu_{\rm HerS}} ( f^{250-\nu_{\rm ACT}} \cdot P^{\nu_{\rm HerS}}_b )/(\sigma_b^{\nu_{\rm HerS}})^2 }{ \sum_{\nu_{\rm HerS}} 1/(\sigma_b^{\nu_{\rm HerS}})^2},
\end{equation}
where $\nu_{\rm ACT}$ is either 148 or 220 GHz, $f^{\nu_{\rm HerS}-\nu_{\rm ACT}} = S(\nu_{\rm ACT})/S(\nu_{\rm HerS})$ is the ratio of the flux density between the different HerS and ACT bands, which is determined by the dust SED, $P^{\nu_{\rm HerS}}_b$ is the value of the HerS stack in bin $b$, the variance for the HerS stack in bin $b$ is $(\sigma_b^{\nu_{\rm HerS}})^2$, and the summation is over the three Herschel bands.  After we normalize the profile, we report the total dust emission at 220 GHz with a parameter $A_{\rm dust}^{220}$.

For the dust SED, we fit a single graybody SED per richness bin,
\begin{equation}
  \label{eq:dustSED}
  \centering
  S(\nu) = A_{\rm dust}  \bigg(\frac{\nu(1+z)}{\nu_0}\bigg)^{\beta_{\rm dust}} \frac{\big(\nu(1+z)\big)^{3} }{{\exp({h \nu (1+z) / k T_{\rm dust}}) - 1}},
\end{equation}
for an overall amplitude, $A_{\rm dust}$, and dust temperature, $T_{\rm dust}$, using the total flux contribution of all the sources within a fixed angular distance from the clusters. 

From the HerS survey source catalog \citep{2014ApJS..210...22V}, we use the summed source flux within an 11$^\prime$-radius aperture around our clusters.  The Herschel source fluxes let us infer $S(\nu)$, and scale the Herschel map stacks down to the ACT frequencies. 
After scaling the Herschel stacks to the ACT bands, we deconvolve the Herschel beams and reconvolve with the appropriate ACT beams.
We fix the emissivity spectral index $\beta_{\rm dust}$ to 1.5, $\nu_0$ to 100 GHz, and the redshift to the average redshift for each of the richness bins. 
\cite{2014A&A...561A..86M} find that setting $\beta_{\rm dust}$ to 1.5 is a good estimate when fitting spectra without enough information to constrain it, and note that it may cause $T_{\rm dust}$ to be slightly overestimated. 
Furthermore, when fitting for a stacked SZ plus graybody spectrum for Planck galaxy clusters, \cite{2018MNRAS.476.3360E} found that the choice of $\beta_{\rm dust}$ had little influence on the measured SZ signal, and use $\beta_{\rm dust}$ = 1.5 to obtain their main results. 
We ran our pipeline on the data from the highest richness bin using $\beta_{\rm dust}$ = 1.4 and 1.6 to see how $\beta_{\rm dust}$ affects our results. We found that varying $\beta_{\rm dust}$ did not significantly affect the mass estimates; the differences were less than 0.1$\sigma$. 
We report results with SED fitting to HerS sources with an 11$^\prime$ aperture in Table \ref{table:mcmcfitparam} to take into account all sources near the clusters, but the results are not sensitive to the precise aperture used.

\subsection{Radio Source Contamination}
\label{sec:radioprof}
Radio sources have been found to reside preferentially in clusters of galaxies and are often associated with emission from the cluster member galaxies \citep{2002ApJ...580...36H,2007ApJS..170...71L,2007AJ....134..897C,2009ApJ...694..992L, 2011ApJ...734..103G}. 
Similarly to the process of measuring dust emission, we look for radio sources near our cluster centers and model their emission at 148 and 220 GHz. 
We use sources from the NVSS survey at 1.4 GHz \citep{1998AJ....115.1693C}. 
Our model for synchrotron emission is:
\begin{equation}
  \label{eq:radioSED}
  \centering
  P^{\rm synch, \nu_{\rm ACT}} = A^{220}_{\rm synch} \cdot P^{1.4}_{\rm synch} \cdot \left(\frac{\nu_{\rm ACT}}{220 \ \rm GHz}\right)^{\alpha_{\rm synch}},  
\end{equation}
where $A^{220}_{\rm synch}$ is an amplitude at 220 GHz, $\alpha_{\rm synch}$ is the spectral index which determines the frequency scaling, and $P^{1.4}_{\rm synch}$ is the normalized stacked synchrotron profile.  Note that the profile alone is determined by 1.4 GHz data, while the amplitude is fit from 148 and 220 GHz ACT data.  The relevant range for the frequency scaling is between those ACT bands. 

The 1.4 GHz synchrotron profile $P^{1.4}_{\rm synch}$ comes from summing the flux density of sources from the NVSS catalog into the bins used for all the stacking in this paper, and dividing the resulting profile by the solid angle in each bin. 
The profile is then normalized to unit integral over solid angle so that $A^{220}_{\rm synch}$ has dimensions of flux density.
When being compared to the stacks at 148 and 220 GHz, the model synchrotron profile is filtered with the same filter used in the analysis, and smoothed by the beam for the appropriate ACT frequency.
Our data cannot constrain the spectral index, $\alpha_{\rm synch}$, so we apply a prior to our MCMC fitting procedure, which is based on ACT and Planck measurements.
When using ACT data and fitting AGN for synchrotron, SZ, and IR emission, \cite{2014MNRAS.445..460G} measured $\alpha_{\rm synch} = -0.55 \pm 0.03$.
For a sample of DSFG's and AGN, \cite{2014MNRAS.439.1556M} find a spectral index between 148 and 218 GHZ of $\alpha^{148-218}_{\rm synch} = -0.55 \pm 0.60$.
\cite{2011ApJ...731..100M} find $\alpha^{20-148}_{\rm synch} = -0.39 \pm 0.04$ and $\alpha^{5-148}_{\rm synch} = -0.20 \pm 0.03$.
There have been several Planck studies measuring the spectral index of radio sources.
For several classes of radio sources, $\alpha_{\rm synch}$ was measured to range between $\sim$ $-$0.37 and $-$0.78  \citep{2016A&A...596A.106P}, and when scaling from 30 GHz, the spectral indices for extragalactic sources were measured to be $-$0.39 and $-$0.37 when scaling to 143 and 217 GHz respectively \citep{2011A&A...536A..12P}.
Given this information, we have placed a prior on $\alpha_{\rm synch}$ that is a Gaussian distribution with a mean of $-$0.5. 
We fixed the prior standard deviation to 0.2, and then tested how adjusting the width of the distribution affects our SZ mass estimates.
Adjusting the prior's standard deviation to 0.1, 0.4, and 0.6 resulted in masses that were within 0.05$\sigma$ of our original mass estimate, and did not affect the uncertainty in the estimate.

\section{Results} \label{sec:results}

We use a Markov-Chain Monte Carlo method to fit the stacked profiles and infer $M_{500}$. We then use richness information available from the IR and optical data to compare to other works.


\subsection{MCMC Fitting}

\begin{table*}
   
  \centering
  \caption{Best-fit parameters for fitting an SZ profile in two ways: correcting for dust and synchrotron emission, and neglecting the dust and synchrotron correction.}
  \begin{threeparttable}
  \begin{tabular}{|*{10}{c|}}
    \hline
    & & \multicolumn{2}{|c}{$10 \leq \lambda < 20$} & & \multicolumn{2}{|c|}{$20 \leq \lambda < 30$} & & \multicolumn{2}{|c|}{$\lambda \geq 30$}\\ \hline
    
    & & Corr. & No corr. & \ &  Corr. & No corr. & \ &  Corr. & No corr. \\ \hline
    
    $M_{500}$ (10$^{13}$ M$_{\odot})$ & & < 1.9 & < 1.1 & \ & < 4.4 & < 3.1 & \ & $8.7^{+1.7}_{-1.3}$ & $6.4^{+1.3}_{-1.3}$ \\ \hline
     
    $p_0^{148} \ (\rm Jy/Sr)$ & & $-20 \pm 40$ & $-10^{+45}_{-50}$ & \ & $-10^{+75}_{-80}$ & $-10^{+75}_{-80}$ & \ & $-100 \pm 130$ & $-90^{+130}_{-150}$ \\ \hline
    
    $p_0^{220} \ (\rm Jy/Sr)$ & & $50 \pm 70$ & - & \ & $-140^{+130}_{-140}$ & - & \ & $7 \pm 220$ & - \\ \hline
    
    $T_{\rm dust} \ (\rm K)$ & & $29.1^{+0.1}_{-0.1}$  & - & \ & $28.0^{+0.1}_{-0.1}$ & - & \ & $27.0^{+0.2}_{-0.2}$ & - \\ \hline 

    $A^{220}_{\rm dust} \ (\rm mJy)$ & & $0.264 \pm 0.001$ & - & \ & $0.268 \pm 0.002$ & - & \ & $0.350 \pm 0.003$ & - \\ \hline
    
    $A^{220}_{\rm synch} \ (\rm mJy)$ & & < 0.5 & - & \ & < 0.5 & - & \ & < 1.5 & - \\ \hline
    
  \end{tabular}
  \begin{tablenotes}
	\item Resulting fit parameters for the three richness bins. 
	The first column for each richness bin shows results from fitting for dust and synchrotron contamination simultaneously with the SZ profile, fixed at the average redshift for each sample, assuming no mass bias. 
	The second column lists the results if we neglect the dust and synchrotron correction and fit an SZ profile directly to the raw, stacked 148 GHz profile. 
  \end{tablenotes}
  \end{threeparttable}
\label{table:mcmcfitparam}
\end{table*}

\begin{figure}
  \centering
  \includegraphics[width=0.5\textwidth]{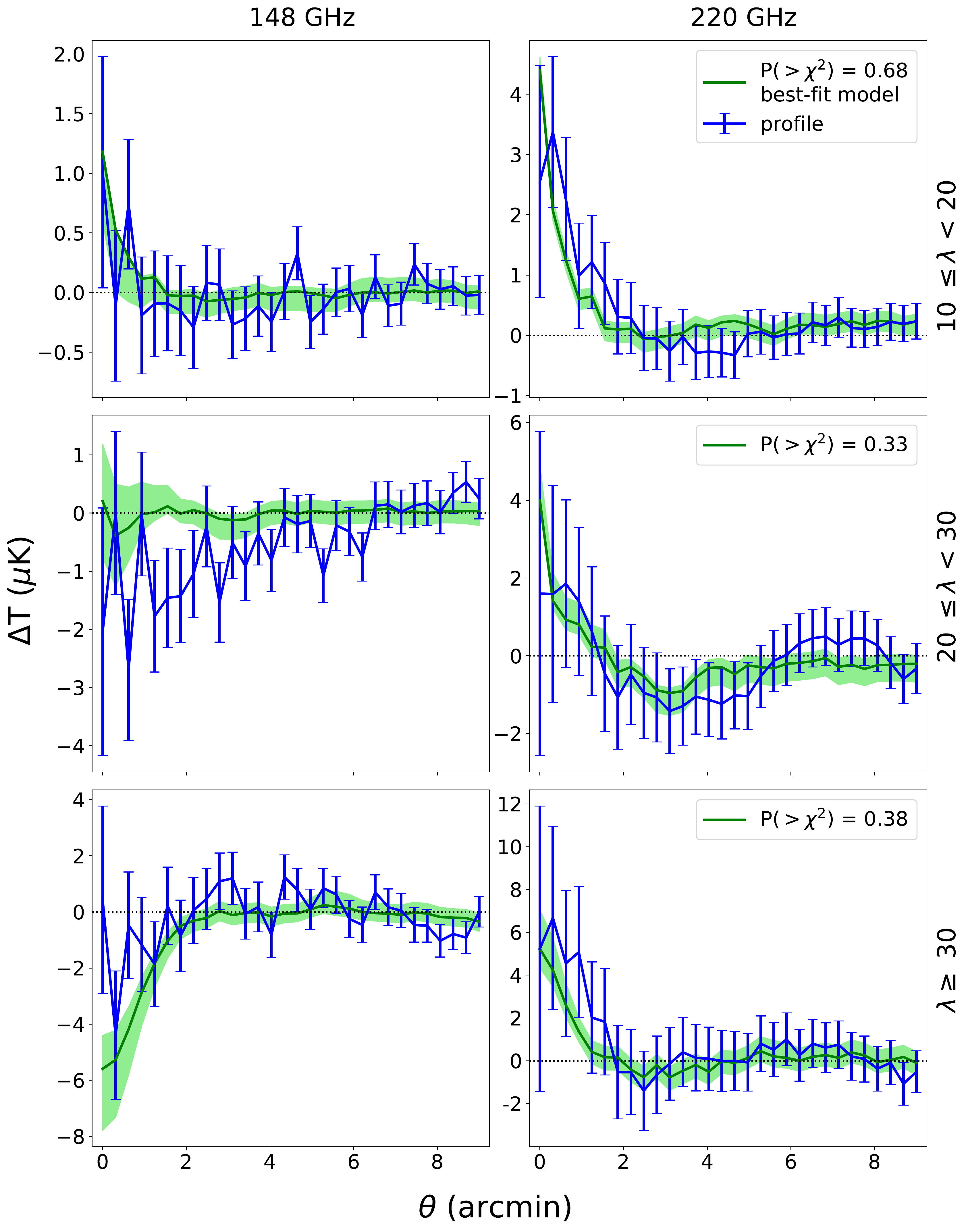}
  \caption{Results at 148 GHz (left) and 220 GHz (right) from fitting for SZ  and contaminating emission for the three richness bins. The blue line is the stacked profile. The green line is the most-likely profile for the combined SZ, dust, and synchrotron model, based on the MCMC chains. The lighter green band bounds the models in the chains between the 16th and 84th percentiles in each angular bin. The legends display the probability to exceed $\chi^2$ for the data minus the model, which takes into account the profiles and correlations at 148 and 220 GHz, as well as their cross-correlations. The PTE's show that all are reasonable fits.}
  \label{fig:mcmcprof}
\end{figure}


\begin{figure}
    \centering
    \includegraphics[width=\columnwidth]{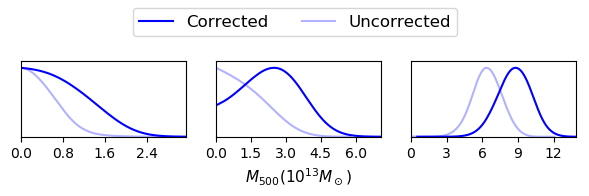}   
    \begin{tabular}{p{2cm} p{2cm} p{2cm}}
         $10 \leq \lambda < 20$ & $\ \ \ \ 20 \leq \lambda < 30$ & $\ \ \ \ \ \ \ \ \ \ \ \ \ \lambda \geq 30$ \\
    \end{tabular}
    \caption{Mass distributions for the three richness bins when correcting for (dark blue)---or not correcting for (light blue)---dust and synchrotron contamination. We see no significant detection in the lowest richness bin. Accounting for contamination slightly increases the estimated masses for the higher richness bins.}
    \label{fig:mcmcmasses}
\end{figure}

\begin{figure*}
  \centering
  \includegraphics[width= 0.9\textwidth]{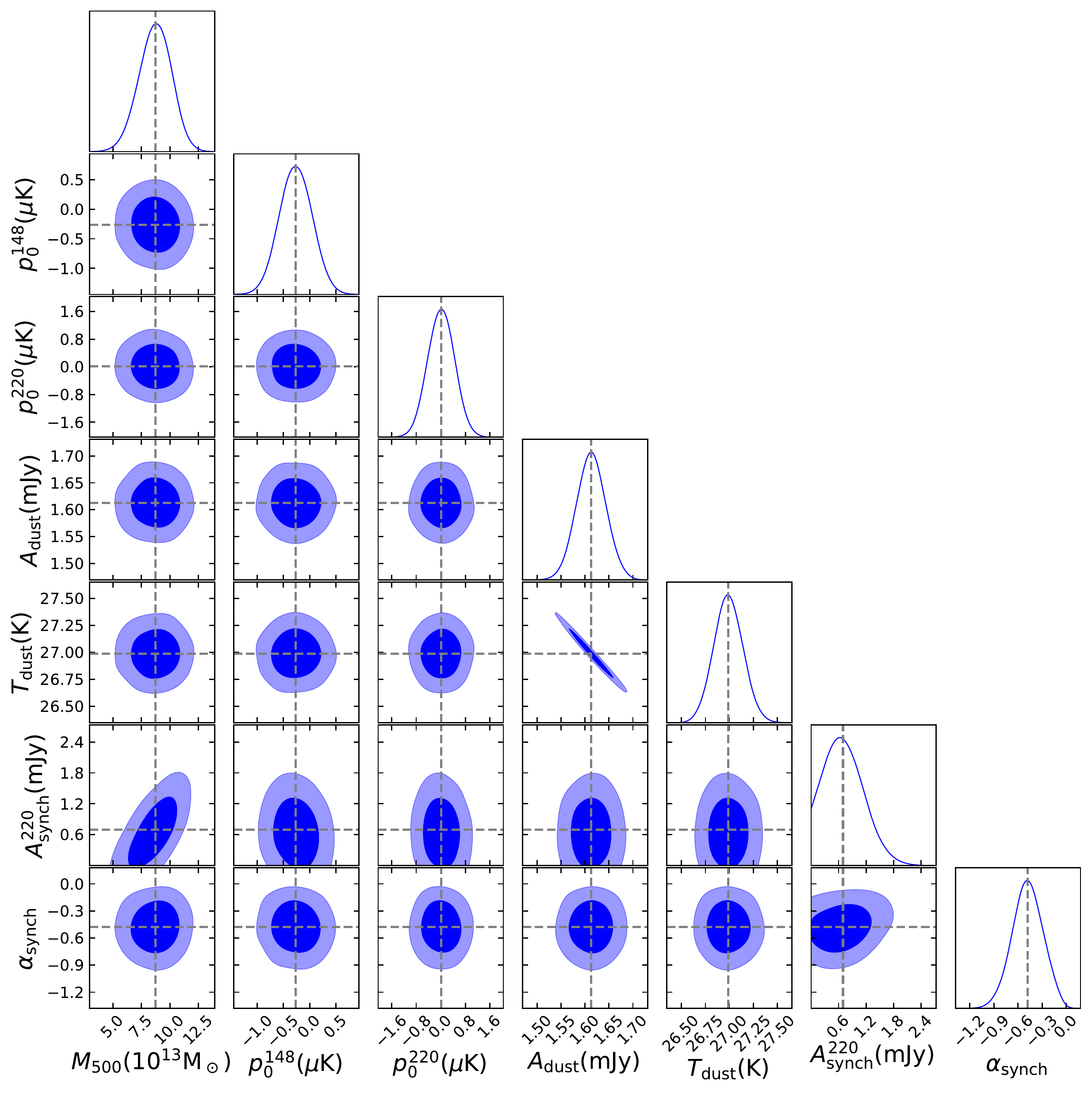}
  
  \caption{Parameter contours from MCMC chains for the highest richness bin, $\lambda \geq 30$. The gray dashed lines pass through the mean value from the probability distributions for each parameter. The dust amplitude and temperature are anti-correlated in all richness bins.}  \label{fig:mcmccontours}
\end{figure*}

To take into account the corrections in the mass fitting, we simultaneously fit for SZ, dust, and synchrotron contributions to the stacked profiles. 
We use a Gaussian likelihood and the affine-invariant Markov Chain Monte Carlo code \textit{emcee} \citep{2013PASP..125..306F}. 
The fit parameters are the average cluster mass $M_{500}$, overall DC offsets for the stacked profiles at 148 GHz and 220 GHz, $p_0^{148}$ and $p_0^{220}$, an amplitude for the graybody SED $A_{\rm dust}$, a dust temperature $T_{\rm dust}$, an amplitude for the synchrotron emission at 220 GHz $A_{\rm synch}^{220}$, and a spectral index for the synchrotron scaling $\alpha_{\rm synch}$. 
We fix the redshift dependence of the dust spectrum (Equation 10) and SZ signal (Equation 4, and calculations of $R_{500}$ and $\rho_c$) to the average redshift for each sample, $z = 0.80, 0.73,$ and 0.70, for the lowest to highest richness bins. 
We apply flat priors which enforce that $M_{500}$, $A_{\rm dust}$, $T_{\rm dust}$, and $A_{\rm synch}^{220}$ are positive, and place no prior on offsets $p_0^{148}$ and $p_0^{220}$.
The prior for $\alpha_{\rm synch}$ is a Gaussian centered on -0.5, with a standard deviation of 0.2, as discussed in Section \ref{sec:radioprof}. 

For each step in the sampler, the sampled mass is used to calculate the SZ signal $y(\theta)$ (Section \ref{sec:methods}), which is convolved with the ACT beam and translated into a temperature profile, $\Delta T(\theta)$.
The sampled dust parameters, $A_{\rm dust}$ and $T_{\rm dust}$ are used to calculate a dust SED and compared to the HerS flux densities, and  determine how to scale the HerS profiles to 148 and 220 GHz. 
The shape of the synchrotron profile is used with the sampled values for $A_{\rm synch}^{220}$ and $\alpha_{\rm synch}$ to estimate the synchrotron contribution at 148 and 220 GHz.

In the likelihood, the sum of the SZ, dust, synchrotron, and DC offset signals are compared to the stacked profile at 148 GHz. 
The sum of the dust, synchrotron, and DC offset signals are also compared to the stacked profile at 220 GHz.
The sampled dust SED is compared to the mean source flux densities for each of the HerS bands.
The  log-likelihood has two parts, one that compares the HERS source fluxes that determine the dust SEDs and one that compares the ACT profiles to the model profiles.
The log-likelihood is:
\begin{equation}
\begin{split}
   -2 \ln \mathcal{L} = & \ (S_{\rm HerS} - S_{\rm model})^{\rm T} \  C_{\rm HerS}^{-1} \ (S_{\rm HerS} - S_{\rm model}) \\
     & + \begin{bmatrix}
     P^{148}_{\rm stack} - P^{148}_{\rm model} \\[1em] P^{220}_{\rm stack} - P^{220}_{\rm model}    
  \end{bmatrix}^{\rm T}
 C_{\rm ACT}^{-1}
  \begin{bmatrix}
     P^{148}_{\rm stack} - P^{148}_{\rm model} \\[1em] P^{220}_{\rm stack} - P^{220}_{\rm model} 
  \end{bmatrix} 
\end{split}
\end{equation}
where $P^{148}_{\rm stack}$ is the stacked profile at 148 GHz, $P^{220}_{\rm stack}$ is the stacked profile at 220 GHz, and $S_{\rm HerS}$ is the mean flux density of sources in the three Herschel bands.
$P^{148}_{\rm model}$ is the model from Equation \ref{eq:prof148}, $P^{220}_{\rm model}$ is the model from Equation \ref{eq:prof220}, and $S_{\rm model}$ is the model from Equation \ref{eq:dustSED}.
$C_{\rm ACT}$ is the full covariance between the 148 and 220 GHz stacks:
\begin{equation}
C_{\rm ACT} = 
      \begin{bmatrix}
    C_{148} & C_{148x220} \\
    C_{148x220}^{\rm T} & C_{220} \\
  \end{bmatrix}.
\end{equation}
$C_{148}$, $C_{220}$, and $C_{148x220}$ include the covariance from stacking on ACT simulations (Figure \ref{fig:covariance}).
To also account for the contribution from the other components to the full covariance, we scaled the covariances for the dust and synchrotron stacked profiles by their appropriate SEDs (Equations \ref{eq:dustSED} and \ref{eq:radioSED}) added them to the ACT-simulation covariance matrices.
$C_{\rm HerS}$ is a diagonal matrix that contains the variance in the mean flux density for each of the three Herschel bands.

The SZ signal does not scale linearly with mass, and choosing one mass value to compare with our stacked profiles may cause us to infer a value for $M_{500}$ that is not characteristic of the clusters in the sample.
To address this, we tested a second fitting method that uses a weighted average of SZ profiles to compare with our stacked profile.
We start with the richness distribution of the clusters in each bin, and use the mass-richness relation of \cite{2015MNRAS.454.2305S} to translate the richness distribution into a mass distribution. 
For each mass $M_{500}^{\rm MCMC}$ that is sampled in the MCMC, we shift the mass probability density distribution to have a mean which is $M_{500}^{\rm MCMC}$, and scale the probabilities accordingly. 
Then we perform a weighted average of the SZ signal with a range of masses, where the weights are the probabilities from the new mass probability density distribution.
The masses that we infer from this fitting method differ from the masses inferred in the main analysis by less than 0.1$\sigma$, but this correction may be more important with future, higher S/N data.

\subsection{MCMC Results}
Figures \ref{fig:mcmcprof}, \ref{fig:mcmcmasses}, and \ref{fig:mcmccontours}  show the results from fitting the stacked profiles for SZ, dust, and synchrotron components. 
Figure \ref{fig:mcmcprof} shows the stacked profiles for the ACT bands in blue, as well as the most-likely models from the MCMC chains in green. 
The light green area encompasses 68\% of the models from the MCMC chains in each annular bin.  
We report the probability to exceed $\chi^2$ (PTE) for data minus model for the best fitting model in each richness bin.  These are computed jointly from the pair of stacks at 148 and 220 GHz and the full covariance matrix.
We calculate the PTE's of the models using 53 degrees of freedom, as there are 60 points in the two stacked profiles and 7 model parameters.

In the lowest richness bin, the best-fitting model primarily shows dust emission and little evidence for an SZ signal. 

In the middle richness bin, there is a mild preference for an SZ signal, but at 148 GHz it is canceled by the source emission.
The middle richness bin also shows a notable, broad decrement in the 148 GHz profile. That decrement is too broad to fit successfully with an SZ profile, and we have tested that increasing the mass worsens $\chi^2$ compared to the best-fit model.  Large scale CMB fluctuations are a possible cause of this signal. 
The CMB is not part of the model, but it is accounted for in the covariance matrix (Section \ref{sec:methods}), and could explain why the PTE shows that this model is a reasonable fit to the data, despite the broad decrement.

In the highest richness bin, we see a clear SZ detection at 148 GHz.
For this case, Figure \ref{fig:mcmccontours} shows all the parameter contours for $M_{500}$, DC offsets in the 148 GHz and 220 GHz stacks, a dust spectrum amplitude, a dust temperature, a synchrotron spectrum amplitude, and the scaling for the synchrotron spectrum from the MCMC chains. 
The synchrotron amplitude and SZ mass are correlated, and the dust parameters are anti-correlated.
The synchrotron amplitude runs into the lower prior of zero in this highest richness bin, and is consistent with zero for the lower richness bins.

The parameter contours for the other richness bins are nearly identical, the main difference being the distributions of the SZ masses.
The separate mass distributions for the three richness bins, correcting for (and neglecting to correct for) contaminating emission, are shown in Figure \ref{fig:mcmcmasses}.
The mass distributions in the lower richness bins run into the prior limit of zero mass, whether correcting or not correcting for contamination.
Accounting for contamination does allow a low significance measurement of the SZ mass for the middle richness bin.
For the higher richness bin, we make a significant mass detection in both cases, and accounting for contamination increases the mass.
The fit parameters are summarized in Table \ref{table:mcmcfitparam}, which show results from simultaneously fitting dust, synchrotron, and SZ components to the ACT data, as well as fitting an SZ profile directly to the data while neglecting to correct for dust and synchrotron emission. 
In the lower richness bins we report 95\% upper limits, as we do not make significant mass estimates.
For the highest richness bin, we can use the maximum likelihood values for the fit parameters in Table \ref{table:mcmcfitparam} with SED equations \ref{eq:dustSED} and \ref{eq:radioSED} to calculate the ratio in the flux between 148 GHz and the NVSS/Herschel bands, finding that $S^{\rm synch}_{148} = 0.097 \cdot S_{1.4}$, $S^{\rm dust}_{148} = 0.015 \cdot S_{500}$, $S^{\rm dust}_{148} = 0.007 \cdot S_{350}$, and $S^{\rm dust}_{148} = 0.005 \cdot S_{250}$.  (Note that the 1.4 GHz NVSS extrapolation may not be reliable because our 148--220 GHz synchrotron spectral index may not be valid down to 1.4 GHz.)

We report an amplitude at 220 GHz for the dust signal as 
\begin{equation}A^{220}_{\rm dust} \equiv \frac{\int d\Omega \ b^{220}(\theta) P^{220}_{\rm dust}(\theta)}{\int d\Omega \ (b^{220}(\theta))^2}, \end{equation}
which is computed by integrating the best-fitting dust profile at 220 GHz $P^{220}_{\rm dust}(\theta)$ over the solid angle $d\Omega$ while weighting by the beam profile $b^{220}(\theta)$.  We see significant dust emission for each of the three richness bins.  The lowest and middle richness bins show similar amounts of dust emission, even though the middle bin's mean richness is 70 percent higher than the lowest bin's. The richest bin has 30 percent more dust emission than the middle bin but 60 percent higher mean richness.

When accounting for dust and synchrotron emission, in the lowest richness bin we place a 95\% upper limit on the mass of $M_{500} \leq 1.2 \times 10^{13}\ \rm M_{\odot}$. 
In the middle richness bin, $M_{500} \leq 4.4 \times 10^{13}\ \rm M_{\odot}$. 
In the highest richness bin, we estimate an SZ mass of $8.7^{+1.7}_{-1.3} \times 10^{13}\ \rm M_{\odot}$. 
Neglecting to account for dust and synchrotron emission decreases the mass by 26\% in the highest richness bin.

As a robustness check, we use the random stacks from Figure \ref{fig:randstacks} to test what mass our pipeline measures when there is no signal. 
Fitting without a dust and synchrotron correction results in probabilities of measuring a mass that pushes up against the lower limit of the prior (zero mass). 
At 95\% confidence, the upper limit of the null mass distributions are $1.9 \times 10^{13} \rm M_{\odot}$, $2.7 \times 10^{13} \rm M_{\odot}$, and $4.8 \times 10^{13} \rm M_{\odot}$, for the lowest to highest richness bins.
As another robustness check, we allow for a negative mass.
We calculate $P^{\rm SZ}$ using the absolute value of the sampled mass, and multiply the profile by -1 if the mass is below zero.
In this case, we find that 50\% of the sampled masses are negative for the lowest richness bin, 65\% for the middle richness bin, and 26\% for the highest richness bin.

\subsection{Impact of Mass Bias}
The relation between SZ signal and cluster mass, $Y$--$M$, is often measured using X-ray derived cluster masses \citep{2010A&A...517A..92A, 2011ApJ...738...48A}, but there are several effects that cause these X-ray mass estimates to be biased low. 
Clusters are assumed to be in hydrostatic equilibrium (HE), but non-thermal pressure support from turbulence and random motions move the clusters away from perfect HE. 
The bias between the true mass and the mass measured by the SZ effect is quantified as $1-b = M_{\rm SZ}/M_{\rm true}$. 
The value for $1-b$ needs to be determined depending on the mass proxy and survey and could lie in a large range \citep{2014A&A...571A..16P}.
In \cite{2014A&A...571A..16P}, $1-b$ is fixed at 0.8. 
\cite{2016JCAP...08..013B} measured the mass bias for high-signal-to-noise clusters from the ACT equatorial survey, using weak-lensing data from the Canada-France-Hawaii telescope stripe 82 survey. 
They found that $1-b$ is 0.98 $\pm$ 0.28 and 0.87 $\pm$ 0.27 when fitting the weak-lensing mass using models based on simulations and an NFW profile, respectively. 
\cite{2019ApJ...875...63M} present the amount of mass bias present in ACTPol clusters when comparing their SZ masses to weak lensing masses derived from Hyper-Suprime Cam data. They find $1-b = 0.74^{+0.13}_{-0.12}$.
There have been several other measurements of $1-b$, with values ranging from 0.58 to 0.95 \citep{2014MNRAS.443.1973V,2015MNRAS.449..685H,2016MNRAS.456L..74S,2016A&A...594A..24P}. 
When comparing our measured $M_{SZ}$ to cluster mass scaling relations, we use the values: $1-b = 1$, 0.8, and 0.6. 
We choose these values to sample the range of values that have been measured, in order to demonstrate how different amounts of mass bias could affect our mass estimates.

\subsection{Richness to Mass Scaling}
\begin{figure}
  \centering
    \includegraphics[width=\columnwidth] {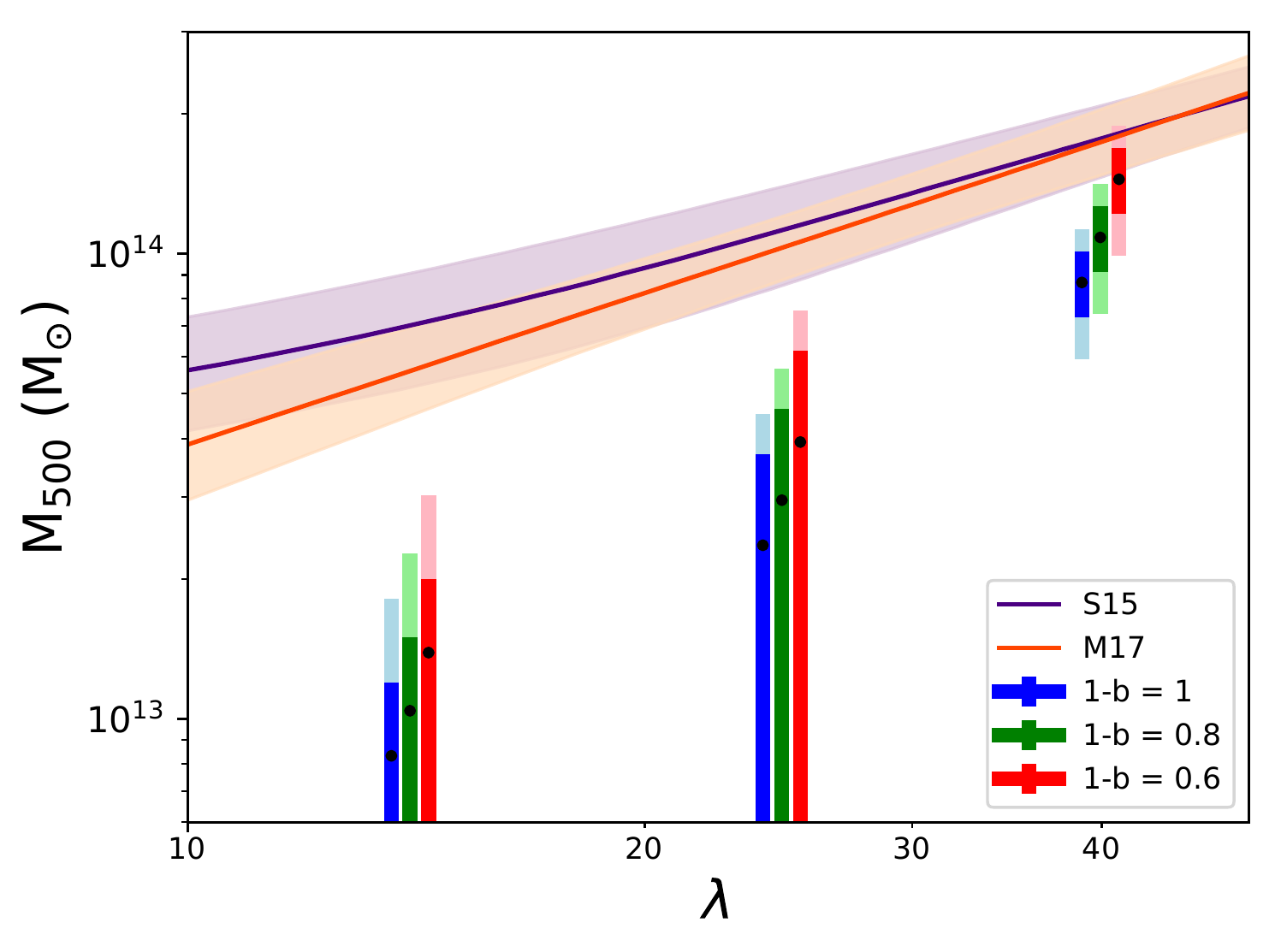}
  \caption{Comparison of SHELA SZ masses with redMaPPer mass-richness relations. The data points show the SHELA clusters in each richness bin and use various values for the mass bias, $1-b$. The data point for the case of $1-b=1$ is plotted at the average richness in each bin. The other two points are slightly offset from the average richness for the sake of clarity. The error bars highlight 68\% and 95\% of the likelihood for mass.  The orange line is the mass-richness relation of \protect \cite{2017MNRAS.469.4899M} which was calibrated using weak-lensing masses of redMaPPer clusters in DES data. The purple line is the mass-richness relation of \protect \cite{2015MNRAS.454.2305S} which was calibrated by abundance-matching SPT clusters that have redMaPPer counterparts. For both models, the average redshift of all the clusters, z = 0.78, is used.}
  \label{fig:y500vslambda}
\end{figure}

We find that the SZ masses and limits we measure are 2--4 smaller than those predicted by optical richness scaling relations. 
If the scalings from the literature held, we should have measured the masses at higher signal-to-noise in all three bins.
Accurately measuring the masses of cluster halos is necessary for clusters to be used for cosmology, making it vital to map out the relation between mass and the cluster observables, such as the SZ signal and richness. 
It is important to compare different observable to mass relations which have different biases and contaminating factors to test if predicted scaling relations hold. 
Multiple studies have seen that the SZ observable has some features that cause a lower SZ signal than predicted when $\lambda$--$M$ relations are extrapolated to lower mass objects \citep{2011A&A...536A..12P,2012PhRvD..85b3005D,2013ApJ...767...38S, 2017MNRAS.468.3347S}, and we explore that possibility here.  

For example, \cite{2017MNRAS.468.3347S} [S17] used a sample of DES redMaPPer clusters and measured their SZ signal by stacking in SPT (South Pole Telescope) data in multiple richness bins.  
They used their own richness-mass model from \cite{2015MNRAS.454.2305S} [S15], which was derived by matching a sample of clusters which were SPT detections with redMaPPer counterparts.  S17 inverted the S15 relation to obtain a mass-richness model.
S17 ran two cases to relate the mass to the SZ signal, using their own $Y_{500} - M_{500}$ model and using the model from A10. 
They compared the $Y_{500}$ expected in each richness bin to what was measured from stacking on the SPT SZ map. 
They did not correct for dust or synchrotron contamination, but discuss how much contamination and mass bias would be necessary to reconcile their measurements.
They found that for clusters with $\lambda$ > 80, $Y_{500}$ is consistent with their model and smaller by 0.61$\pm$0.12 from A10. 
For 20<$\lambda$<80, they found that the SZ signal was smaller by a factor of ~0.2--0.8, with higher richness bins and the S15 $Y_{500} - M_{500}$ showing better agreement. 
They discuss possible explanations for this, such as a richness-dependent bias caused by contamination of the SZ and richness observables. 
Another possibility they discuss is a bias in estimated halo mass. 
This could be caused by a contamination in the richness from line of sight projections, contamination of the SZ observable by dust or synchrotron emission, a larger offset in SZ-optical centering than accounted for, or a larger intrinsic scatter in richness-mass relation at lower richness.

In Figure \ref{fig:y500vslambda} we compare our results to two mass-richness relations.
First, we compare with the $M_{500}$-$\lambda$ relation of S15. 
To make this comparison, we need to take into account that they have modeled $P(\lambda|M_{500})$ as opposed to $P(M_{500}|\lambda)$. 
Their model is a log-normal distribution with mean:
\begin{equation}
\begin{split}
\langle \ln\lambda | M_{500},z\rangle = & \ln A_{\lambda} + B_{\lambda} \ln\bigg(\frac{M_{500}}{3 \times 10^{14}h^{-1}\rm M_{\odot}}\bigg) \\ & + C_{\lambda}\ln\bigg(\frac{E(z)}{E(z=0.6)}\bigg)
\end{split}
\end{equation}
To invert this relation, we perform the following operation:
\begin{equation}
P(M_{500}|\lambda^{\rm obs}) \propto P(M_{500},z) \int P(\lambda^{\rm obs}|\lambda) \  P(\lambda|M_{500},z) \ d\lambda,
\end{equation}
where $P(\lambda|M_{500},z)$ is the SPT probability marginalized over the fit parameters $A_{\lambda}$, $B_{\lambda}$, and $C_{\lambda}$. 
$P(\lambda^{\rm obs}|\lambda)$ is the probability for observing a value for the richness given the true richness, and $P(M_{500},z)$ is proportional to the halo mass function. 
This inverted relationship is used in Figure \ref{fig:y500vslambda} to compare against our SZ masses.

\cite{2017MNRAS.469.4899M} [M17] use weak lensing to measure $P(M_{200}|\lambda)$ for redMaPPer clusters in DES. 
For comparison, we translate their model in terms of $M_{200m}$ to $M_{500c}$, where $m$ denotes the density contrast relative to the mean matter density and $c$ is relative to the critical density at that redshift. 
This relation is plotted in Figure \ref{fig:y500vslambda}. 
Although we do not compare our data to other redMaPPer richness-mass models, we note that there is good agreement between relations from S15, \citet[][which is calibrated for clusters in SDSS]{2017MNRAS.466.3103S}, and \citet[][which is calibrated for clusters in DES Year 1 data]{2019MNRAS.482.1352M}.

We specifically compare the mass--richness models from S15 (as they used a sample of redMaPPer clusters which were identified in SZ data) and M17 (as their redMaPPer richnesses are from DECam imaging, like the SHELA data). 
We do not expect the higher redshift range of our clusters to affect this comparison as there is no significant redshift evolution in either model. 
We plot our data against these models in Figure \ref{fig:y500vslambda}. 
For the SHELA sample, the $M_{500}$ from SZ profile fitting and the average richness per bin are used. 
We tested different ways to represent our richness bins, such as using mass-weighted and SZ-weighted average richnesses, but find that they are similar to the mean richness per bin, so we simply use the mean value. 
The average richnesses for the bins are approximately 14, 24, and 39. 
Similarly to the results of S17, we find that the SZ decrements of our clusters indicate that they are less massive than predicted by their richnesses and the cluster mass-richness relationships.
Without accounting for mass bias for the highest richness bin, the predicted masses from S15 and M17 are 2.0$\pm$0.5 and 1.9$\pm$0.5 times larger than the SZ mass we found. 
With a  mass bias of $1-b = 0.6$, the predicted masses from S15 and M17 are 1.2$\pm$0.3 times larger.
As is shown in Figure \ref{fig:y500vslambda}, a smaller value for $(1-b)$ would be necessary to reconcile our highest richness mass estimate with the richness-based mass models. 
The lower two richness bins have masses that are significantly below the predicted values, even when accounting for mass bias.
There would have to be a significant error in the richness to cause a discrepancy this large, therefore there must be other factors in play. 
Most of the possible explanations coincide with those discussed in S17: more contamination in the SZ signal than we have accounted for, contamination in the richness estimate, or these mass-richness relations are failing when extrapolated to low richness. 
For the SHELA sample specifically, there could also be a differences in the meaning or interpretation of the richness, as the redMaPPer algorithm included IR data in addition to optical data when identifying this sample.


\section{Discussion} \label{sec:conclusions}
We have presented the analysis of stacked SZ profiles for a sample of IR and optically-selected clusters from the Spitzer-HETDEX Exploratory Large Area survey. 
We split this sample into three richness bins: $10 \leq \lambda < 20$, $20 \leq \lambda < 30$ and $\lambda \geq 30$. 
There are 840, 172, and 70 clusters in the richness bins, from lowest to highest. 
At the SZ null (220 GHz), the stacked profiles exhibited an excess signal, which we attributed to dust emission from cluster member galaxies. 
For each bin, we fit for a dust SED using sources from the Herschel Stripe 82 survey catalog to extrapolate the Herschel stacks to 148 and 220 GHZ. 
We also fit for a synchrotron amplitude while setting a prior on the synchrotron spectral index, which we use to estimate the contributions from synchrotron emission at 148 and 220 GHz. 
We fit for an SZ profile using a universal galaxy cluster pressure profile which translated our temperature decrement into a halo mass. 
For each richness bin, we used an MCMC procedure that simultaneously fitted for the SZ, dust, and synchrotron components while fixing the redshift to the average of the cluster sample.
We made a detection of dust emission, and placed upper limits on the synchrotron emission.
We compared the chains with and without the dust and synchrotron correction, and found that for the highest richness bin, neglecting to correct for contamination decreases the estimated mass by 26\%.
In the lower richness bins, we did not make significant SZ mass detections.

The SHELA cluster catalog obtains richnesses from the redMaPPer algorithm. 
We compared our SZ mass estimates and richness data to two models which have mapped out the richness-mass relation for redMaPPer clusters. 
SZ mass measurements of optically-selected clusters are generally smaller than expected when comparing to mass-richness models, and the SHELA clusters follow this trend. 
For all three richness bins, it would take a large amount of mass bias to reconcile our estimates with redMaPPer mass-richness relations. 
For the $\lambda \geq 30$ bin, the masses predicted by mass-richness relations are 2$\pm$0.5 times larger than our SZ mass estimates.
Taking into account a mass bias value of $1-b = 0.6$, the predicted masses are 1.2$\pm$0.3 times larger.
For the lower two richness bins, our SZ mass estimates fall below the predicted masses even when taking this mass bias into account.

This work is a step toward studying characteristics of galaxy clusters over a range of redshifts and masses. 
In the future, wider and deeper coverage by Advanced ACT and the Simons Observatory will advance this study by observing a large number of clusters in five frequency bands, and by increasing overlap with optical surveys such as BOSS, HSC, DES, DESI, and LSST \citep{2016SPIE.9910E..14D}. 
It may be fruitful to repeat this type of analysis for more infrared-selected objects. Similar studies could be done using the Spitzer IRAC Equatorial Survey \citep[SpIES,][]{2016ApJS..225....1T} which is shallower than SHELA but larger, covering an adjacent $\sim$115 square degrees of Stripe 82, or using the MaDCoWS cluster catalog which covers the full extragalactic sky at 0.7 $\lesssim z \lesssim 1.5$ \citep{2019ApJS..240...33G}.

\section*{Acknowledgments}
BJF and KMH acknowledge support by the National Aeronautics \& Space Administration through the University of Central Florida's NASA Florida Space Grant Consortium and Space Florida. KMH also acknowledges support from the U.S. National Science Foundation through award AST-1815887. LK and CP acknowledge support from the NSF through grants AST-1413317 and 1614668. NS acknowledges support from NSF grant number AST-1513618.  This work was also supported by the NSF through awards AST-0408698 and AST-0965625 for the ACT project, as well as awards PHY-0855887 and PHY-1214379. Funding was also provided by Princeton University, the University of Pennsylvania, and a Canada Foundation for Innovation (CFI) award to UBC. ACT operates in the Parque Astron\'omico Atacama in northern Chile under the auspices of the Comisi\'on Nacional de Investigaci\'on Cient\'ifica y Tecnol\'ogica de Chile (CONICYT). Computations were performed on the GPC supercomputer at the SciNet HPC Consortium. SciNet is funded by the CFI under the auspices of Compute Canada, the Government of Ontario, the Ontario Research Fund - Research Excellence; and the University of Toronto. The development of multichroic detectors and lenses was supported by NASA grants NNX13AE56G and NNX14AB58G. We thank our many colleagues from ABS, ALMA, APEX, and Polarbear who have helped us at critical junctures. Colleagues at AstroNorte and RadioSky provide logistical support and keep operations in Chile running smoothly. We also thank the Mishrahi Fund and the Wilkinson Fund for their generous support of the project. The Institute for Gravitation and the Cosmos is supported by the Eberly College of Science and the Office of the Senior Vice President for Research at the Pennsylvania State University. 




\bibliographystyle{mnras}
\bibliography{shela_sz}


\bsp	
\label{lastpage}
\end{document}